\documentclass[12pt]{article}
\usepackage{amsfonts}

\def\one{1\hskip-.37em 1}

\def\half{\textstyle{\frac{1}{2}}}

\def\threebytwo{\textstyle{\frac{3}{2}}}

\def\p{\phi}

\def\l{\lambda}

\def\t{\textstyle}

\def\ra{\rightarrow}
\def\tint{{\textstyle\int}}

\def\hpi{{\hat\pi}}

\def\s{\hskip.08em}

\def\d{\partial}

\def\b{\begin{eqnarray*}}  
\def\e{\end{eqnarray*}}    
\def\bn{\begin{eqnarray}}  
\def\en{\end{eqnarray}}   

\def\hk{\hat{\k}}
\def\hp{\hat{\p}}
\def\<{\langle}

\def\>{\rangle}

\def\no{\nonumber}

\def\k{\kappa}
\def\{{\lbrace}
\def\}{\rbrace}

\title{Solving the Insoluble: \\A New Rule for Quantization}
\author{John R. Klauder\footnote{john.klauder@gmail.com}\\
Department of Physics and Department of Mathematics\\
University of Florida,
Gainesville, FL 32611-8440}
\date{ }
\begin{document}
\maketitle

\begin{abstract}The rules of canonical quantization normally offer good results, but sometimes they fail, e.g., leading  to quantum triviality
($=$ free) for certain examples that are  classically nontrivial ($\ne$ free). A new procedure, called Enhanced Quantization,
relates classical models with their quantum partners differently and leads to satisfactory results for all systems. This paper
features enhanced quantization procedures and provides highlights of two examples, a rotationally symmetric model and an ultralocal scalar
model,  for which canonical quantization fails while enhanced quantization succeeds.  \end{abstract}

 \section{From Canonical Quantization \\to Enhanced Quantization}
 For a single degree of freedom, canonical quantization (CQ) takes classical phase space variables, $(p,q)$, and promotes them into Hilbert space operators, $(P,Q)$. These basic
 operators obey the commutator $[Q,P]=i\hbar \one$, an analog of the classical Poisson bracket $\{q,p\}=1$.
  Additionally, the classical Hamiltonian  $H(p,q)$
  is promoted to the quantum Hamiltonian $ {\cal{H}}(P,Q)$, specifically
 ${{\cal{H}}}(P,Q)=H(P,Q)+{\cal{O}}(\hbar; P, Q)$, with ${\cal{O}}(0; P, Q)\equiv 0$,  provided that the
 phase space variables $p$ and $q$ are ``Cartesian coordinates'' \cite{D} (footnote, page 114), despite the fact that phase space does not have a metric necessary to determine Cartesian variables. In the author's view, it is likely that this important remark by Dirac was relegated to a footnote because he could not justify it any further. Moreover, the operators $P$ and $Q$ need only be Hermitian despite the fact that such
 operators may not be self adjoint and thus have complex eigenvalues (e.g., if $Q>0$ then $P$ can be Hermitian but it cannot be self adjoint). Thus canonical quantization may encounter some difficulties in the classical/quantum transition.

Enhanced quantization (EQ) features a different classical/quantum connection story. It starts by introducing self-adjoint operators $P$ and $Q$ that satisfy the commutator $[Q,P]=i\hbar \one$. The quantum Hamiltonian ${\cal{H}}(P,Q)$ is chosen as a self-adjoint operator built from the basic variables.
Schr\"odinger's equation follows from the action functional    %
\bn A_Q=\tint_0^T \<\psi (t)|[ i\hbar(\d/\d t) - {{\cal{H}}}(P,Q)] |\psi (t)\> \,  dt \en   
 by stationary variations of the normalized vectors $|\psi (t)\>$. However, a {\it macroscopic} observer can only vary a subset of vectors of a {\it microscopic} system such as
 \bn  |p,q\> \equiv e^{-iqP/\hbar}\s e^{ipQ/\hbar} |0\>\;, \en    where      
   $(p,q)\in \mathbb{R}^2$, and the normalized fiducial vector $|0\>$ is chosen as a solution of the equation $(\omega Q +i P ) |0\>=0$.  Recall that Galilean invariance permits a change of the system by position or velocity (note: $v=p/m$ for fixed $m$) by a change of the observer {\it without disturbing the system}. All vectors $|p ,q\>$ are normalized as well because $P$ and $Q$ are self adjoint. The reduced (R) quantum action functional is given by
 \bn A_{Q(R)}=\tint_0^T \<p(t),q(t)|[i\hbar(\d/\d t)-{\cal{H}}(P,Q)]|p(t),q(t)\>  \;  dt \;,   \en
   which leads to
  \bn A_{Q(R)}=\tint_0^T [p(t) \dot{q}(t) - H(p(t),q(t))]\;  dt\;, \en
 a natural, potential, candidate to be the {\it classical Hamiltonian}, with
 $\hbar>0$, {\it as it is in the real world!} In this expression
 \bn  && \hskip-3.4em H(p,q)\equiv  \< p,q| {\cal{H}}(P, Q) |p,q\> \no\\
       &&= \<0| {\cal{H}}(P+p\one, Q+q\one) |0\>  \no \\
      &&= {\cal{H}}(p,q)+{\cal{O}}(\hbar;p,q) \;. \en
   Note well that, apart from possible $\hbar$ corrections, ${\cal{H}}(P,Q)=H(P,Q)$, which {\it exactly}  represents the goal of seeking Cartesian coordinates. Although phase space has no metric to identify Cartesian coordinates, Hilbert space has one. It follows that a scaled form of the Fubini-Study metric \cite{FS} for ray vectors, specifically $d\sigma^2\equiv 2\hbar\s [\s \|\s d|p,q\>\|^2-|\<p,q|\,d|p.q\>|^2]$, yields the result $ d\sigma^2=\omega^{-1} dp^2+\omega\s dq^2$, which   indeed implies Cartesian coordinates for a flat phase space. {\it Consequently, we find that EQ can lead to the same good results given by CQ! But EQ can do even more, and so we claim that $CQ\subset EQ$}.

      What happens if $q>0$, and thus $Q>0$, and $P$ cannot be made self adjoint?
Observe that $i \hbar\s Q =Q[Q,P]= [Q,D]$, where $D\equiv \half
 (QP+PQ)$. Although $P$ cannot be self adjoint, $D$ can be self adjoint and there are two important self-adjoint representations of the affine commutation relation, $[Q,D]=i\hbar\s Q$, one where $Q>0$, the other where $Q<0$. Consider $q>0$, hence $Q>0$,  and a new set of reduced affine ($a$)  states
   \bn |p,q;a\>\equiv e^{ipQ/\hbar} e^{-i \ln(q) D/\hbar} |\tilde{\beta}\>\;, \en
   where $[ ( Q  - 1)+iD/\tilde{\beta} ] |\tilde{\beta}\>=0$. It follows that
  \bn &&A_{Q( R )}=\tint_0^T   \<p(t),q(t); a| [ i\hbar  (\d/\d t)-
 {\cal{H}}'( D, Q) ] |p(t), q(t); a\> \;  dt \no \\
    &&\hskip2.79em =\tint_0^T [-q(t) \dot{p}(t)-H(p(t), q(t)) ]\;   dt \;,  \en
 which also serves as an acceptable classical canonical Hamiltonian! Are the phase-space variables Cartesian? With $q>0$ they cannot be Cartesian. The Fubini-Study  metric confirms that statement with $d\sigma^2=\tilde{\beta}^{-1}\s\s  q^2 \s dp^2 + \tilde{\beta}\s q^{-2}\s dq^2$, which is the metric of a surface of constant negative curvature: $-2/\tilde{\beta}$.

        In EQ the classical and quantum Hamiltonian are related by the weak correspondence principle \cite{DD},
    \bn H(p, q)= \<p, q | {\cal{ H}}(P, Q) | p, q \>\;, \label{e3} \en
 which holds for self-adjoint canonical variables, and in the form
     \bn H(p, q)=H'(pq, q)=\<p,q;a| {\cal{H}}'(D, Q) |p, q ; a\>\;, \label{e4} \en
 for self-adjoint affine variables.
 In CQ the canonical operators $P$ and $Q$ are required to be irreducible, and in EQ they may also be irreducible. However, from (\ref{e3}) and
         (\ref{e4}), it follows that sometimes the
         quantum variables may also be {\it reducible}.  Certain problems can be properly solved with reducible operators that cannot be properly solved with irreducible  operators. This feature may seem foreign to the reader, but perhaps two examples of the power of EQ may help the reader to overcome any doubts.

\section{Rotationally Symmetric Models}
This example involves many variables, and it has been examined before \cite{ka}; here we only offer an overview of the solution. The classical Hamiltonian is given by
\bn H(p, q)=\half \Sigma_{n=1}^N [p_n^2+m_0^2 q_n^2]
+\l_0 \{\Sigma_{n= 1}^N q_n^2 \}^2 \;, \en
where $N\le \infty$, and the Poisson bracket is $\{q_k, p_l\}=\delta_{kl}$. Observe that this model has a symmetry called
``shuffle symmetry". For example, if $N=52$, and the variables for only $n=1$ and  $n=2$ have non-vanishing initial values the results are identical if the same initial data was instead given to $n= 7$ and $n=23$, etc.
     For $\l_0>0$ and  $N=\infty$ the quantum model is trivial (i.e., Gaussian) when CQ and irreducible operators are used (if you have doubts, try to solve it; or see \cite{ka,BCQ} for a proof). To obtain reducible operators we introduce a second  set of canonical operators $\{R_n,S_n\}$
      which commute with all $\{P_n,  Q_n\}$. Triviality of such models forces us to restrict the fiduciary vector to be a Gaussian in a Schr\"odinger representation,
     i.e., where $Q_n\s|x\>=x_n\s |x\>$. Thus $(m_0 Q_n+i P_n) |0\>=0$ for the free model, with the property that
     \bn (m_0 Q_n-i P_n) \cdot(m_0 Q_n+i P_n)&&\hskip-1.3em  =P_n^2+m_0^2\s Q_n^2+i\s m_0\s [Q_n,P_n] \no \\
               &&\hskip-1.25em  =\, :\s P_n^2+m_0^2\s Q_n^2\s :\;, \en
     for all $n$, $1\le n\le N$, with normal ordering denoted by :$\s(\cdot)\s$:.  For the interacting model this
     is changed to $[m(Q_n+\zeta S_n)+i P_n] |\zeta\>=0$ and
$[m(S_n+\zeta Q_n)+ i R_n ] |\zeta\>=0$, for all $n, \, 1\le n\le N$, where $ 0<\zeta< 1$. We next introduce new reduced states as
   \bn |p,q;\zeta\>\equiv e^{-i\Sigma_{n=1}^N q_n P_n/\hbar}\s e^{i\Sigma_{n=1}^N p_n Q_n/\hbar} |\zeta\> \;, \en
which still span the relevant Hilbert space. Then it follows  that
\bn   &&\hskip-3.4em H( p, q)=\<p, q;\zeta| \,\half \, \{: \Sigma_{n=1}^N [P_n^2+m^2(Q_n+\zeta S_n)^2] :\no \\
    &&\hskip6em + : \Sigma_{n=1}^N [R_n^2+m^2(S_n+\zeta Q_ n)^2] : \} \no \\
   &&\hskip6em  + v : \{ \Sigma_ {n=1}^N [R_n^2+m^2(S_n+\zeta Q_n)^2] \}^2 : |p,q;\zeta\> \no \\
    &&=\half \Sigma_{n=1}^N [p_n^2+m^2 (1+\zeta^2) q_n^2] +
      v\s \zeta^4 m^4 \{ \Sigma_{n=1}^N q_n^ 2 \}^2 \no \\
  && \equiv \half \Sigma_{n=1}^N [p_n^2+ m_0^2 q_n^2]
  + \l_0 \{ \Sigma_{n=1}^N q_n^2 \}^2 \en
  as desired!

  Clearly, the energy levels for this quantum Hamiltonian obey shuffle symmetry; however, using the rules of CQ, shuffle symmetry would seem to fail even for
  finite $N$. Indeed, a comparison of complete results predicted for this model by CQ and by EQ, say for $N=3$,  would be of considerable interest.

  Further discussion of these models is available in \cite{ka,BCQ}.

  \section{Ultralocal Models}
 The classical Hamiltonian for this model is given by
 \bn H(\pi,\p)=\tint \{\s\half [\pi(x)^2  +m_0^2\s \p(x)^2\s]+\l_0\s \p(x)^4\s \}\, d^s\!x \;, \label{kp} \en
where $x\in \mathbb{R}^s$, $s\in \{1,2,3,\cdots\}$ and  with the Poisson bracket $\{\p(x), \pi(x')\}=\delta^s(x-x')$. This model has also been discussed before, e.g., \cite{BCQ,ul,UL2}, and we will offer just an
overview of its quantization. Note that there is only a time derivative, e.g., $\dot{\p}(t,x)=\pi(t,x)$, but no space derivatives, a fact which
leads to the name for such models. Given the quantum Hamiltonian ${\cal H}$ for this problem in the Schr\"odinger representation, the ground state
should have the form $\psi_0(\phi)=\exp\{ - \half\tint y[\phi(x);\hbar] \,d^s\!x\s\}$.

Preparatory to additional quantum analysis, let us first introduce an $s$-dimensional  spatial lattice for (\ref{kp}) of step $a>0$ to approximate
the  spatial integral, which leads to the expression for a regularized classical Hamiltonian given by
  \bn H_K(\pi,\p)=\Sigma_k [ \half [\s\pi_k^2+m_0^2\s\p_k^2\s]+\l_0\s \p_k^4 ]\,  a^s\;, \en
  where $k=\{k_1,k_2,\ldots,k_s\}\in K$, $k_j\in\mathbb{Z}$, where $K$ is large but finite  and $a^s$ represents a cell volume. Clearly, a suitable limit of $K\ra \infty$ and $a\ra 0$ such that $K\s a^s\ra V$ where $V\le\infty$ is the spatial volume implicit in (\ref{kp}). Such a limit also applies to the classical equations generated by the classical Hamiltonian.

  Next we discuss a quantization based on CQ for the lattice Hamiltonian prior to taking a spatial limit with $a\ra0$. Since there is no connection of the dynamics between spatial points, it follows that at each lattice site $\pi_k\ra a^{-s}\s P_k$ and $\p_k\ra Q_k$ and $H_k(\pi_k,\p_k)\ra {\cal{H}}_k(P_k,Q_k)$. Moreover,
  at each site ${\cal{H}}_k(P_k,Q_k)=\{\,\half[\s a^{-2s}\s P_k^2+m_0^2\s Q_k^2]+\l_0\s Q_k^4 +{\cal{O}}(\hbar;P_k,Q_k; a)\,\}\, a^s$.
  Here the auxiliary term is assumed to be polynomial in the operators, and it is adjusted so that the ground state $|\psi_{0;k}\>$ fulfills the relation ${\cal{H}}_k(P_k,Q_k)\, |\psi_{0;k}\>=0$.
  The exact normalized Schr\"odinger representation $\<\p_k|\psi_{0;k}\>$ is not known, but, apart from a choice of phase, its shape is roughly of the form $\exp[- Y(\p_k;\hbar,a)\s a^s/2]$, where $-\infty<Y(\p_k;\hbar,a)=Y(-\p_k;\hbar,a)<\infty$. The characteristic function of the regularized overall ground-state distribution is given by
      \bn &&C_K(f)=\Pi_k\,\tint e^{i f_k \p_k\s a^s/\hbar}\,  e^{- Y(\p_k;\hbar,a)\s a^s}\,d\p_k/\tint  e^{- Y(\p_k;\hbar,a)\s a^s}\; d\p_k \no\\
     &&\hskip3.13em =\Pi_k\,\tint e^{i f_k \p_k\s }\,  e^{- Y(\p_k\hbar/ a^s;\hbar,a)\s a^s}\,d\p_k/\tint  e^{- Y(\p_k\hbar/ a^s;\hbar,a)\s a^s}\; d\p_k \;. \en
  The final issue, now, is taking the continuum limit as $a\ra0$. To ensure a reasonable limit it is helpful to expand the term involving $f_k$ in a
  power series, which leads to
     \bn C_K(f)=\Pi_k\{ 1-\t{\frac{1}{2!}} f_k^2\<
     \p_k^2\>+\t{\frac{1}{4!}}f_k^4\<\p_k^4\>-\t{\frac{1}{6!}}f_k^6\<\p_k^6\>-\ldots \}\;. \en
     In order to achieve a meaningful continuum limit, it is necessary that $\<\p_k^2\>\propto a^s$, which, in the present case,
     means that $\<\p_k^{2\s j}\>\propto a^{s\s j}$, leading to the result that the continuum limit becomes
     $C(f)=\exp[-B\tint f(x)^2\,d^s\!x ]$ for some $B$, $0<B<\infty$.
     This result, which is a standard result of the Central Limit Theorem \cite{CLT}, implies a Gaussian (= free) ground state for CQ.

     Now let us examine what EQ has to say about this model. First we introduce classical affine fields $\p(x)$ and $\k(x)\equiv \pi(x)\s \p(x)$,
     with a Poisson bracket given by $\{\p(x),\k(x')\}=\delta^s(x-x')\s\p(x)$. Next the classical Hamiltonian is recast as
       \bn H'(\k,\p)=\tint \{\s\half\s [\s\k(x)\s\p(x)^{-2}\s\k(x) + m_0^2\s\p(x)^2] +\l_0\s \p(x)^4 \} \, d^s\!x \;. \en
       A quantum study begins by promoting the classical affine fields to operators $\k(x)\ra\hk(x)$ and $\p(x)\ra\hp(x)$, $\hp(x)\ne0$, which satisfy
       $[\hp(x),\hk(x')]=i\hbar\s\delta^s(x-x')\s \hp(x)$. Formally, it follows that
       \bn  \hk(x)\s\hp(x)^{-2}\s\hk(x)=\hpi(x)^2+{\t{\frac{3}{4}}}\s\delta^s(0)^2\s \hp(x)^{-2} \;.  \en
       On the same spatial lattice as before, the quantum Hamiltonian is chosen as
       \bn {\cal{H}}'_K(\hk,\hp)=\Sigma_k \{\half [ a^{-2s}\s\hbar^2\s \d^2/\d\p_k^2\s + a^{-2s}\s F\hbar^2 \p_k^{-2} +m_0^2\s
        \p_k^2] +\l_0 \p_k^4\s -E_0\}\,a^s . \en
       where $F\equiv(\half-b\s a^s)(\threebytwo-b \s a^s)$ is a regularized version of $\t{\frac{3}{4}}$, and $b$ is a fixed constant so
        that $b\s a^{s}$ is dimensionless. The ground state $\psi_0(\p)$ satisfies ${\cal{H}}'_K(\hk,\hp)\s \psi_0(\p)=0$ and has a form such that
       \bn |\psi_0(\p)|^2=\Pi_k\,(ba^s)\,e^{-Z(\p_k;\hbar,a)\s a^s}\,|\p_k|^{-(1-2ba^s)}\;, \label{pp} \en
       for $ba^s\ll1$. The term $Z$ is adjusted such that each factor in (\ref{pp}) is normalized, and the term $(ba^s)$ provides normalization
       when the integration passes through $\p_k=0$. Again we study the characteristic function
       \bn &&\hskip-2.5em C(f)=\lim_{a\ra0} \Pi_k\tint\s\{\s e^{if_k\p_k\s a^s/\hbar}\,(ba^s)\,e^{-Z(\p_k;\hbar,a)\s a^s}\,|\p_k|^{-(1-2ba^s)}\;d\p_k \s\}\no\\
           &&=\lim_{a\ra0}\Pi_k\{1-(ba^s)\tint[1-e^{if_k\p_k\s a^s/\hbar}] \,e^{-Z(\p_k;\hbar,a)\s a^s}\,|\p_k|^{-(1-2ba^s)}\;d\p_k \s\}\no\\
           &&=\exp\{-b\tint d^s\!x\,\tint[1-e^{if(x)\s\l/\hbar}]\, e^{-z(\l;\hbar)}\,d\l/|\l| \}\;, \label{uu} \en
           where $\l=\p_k\s a^s$. Observe that using (reducible) affine fields has given rise to the singular factor with the coefficient
           of $\t{\frac{3}{4}}=\half\cdot\threebytwo$ with a value so that, when regularized, the pre-factor $(ba^s)$ serves to provide the proper normalization when $(ba^s)\ll1$ and suitable factors in $Z$ undergo operator product expansion renormalization which leads to $z$.
           Besides a Gaussian distribution found by CQ, the resultant form (\ref{uu}) given by EQ
            is  the only other result of the Central Limit Theorem, namely,  a (generalized) Poisson distribution. Moreover,
           the classical limit as $\hbar\ra0$ for this solution has been shown \cite{E3} to yield the starting classical Hamiltonian.

           As an example, suppose we take the limit $\l_0\ra0$. The resultant characteristic function is given by
              \bn C_0(f)=\exp\{-b\tint d^s\!x\s\tint[1-e^{if(x)\s\l/\hbar}\s]\s e^{-b\s m\l^2/\hbar}\,d\l/|\l|\s \} \;.\en
              Evidently, this expression doesn't imply a ``traditional'' free model; however, the ground state described here reflects the fact that the
              domain of the classical free action functional is strictly larger than the domain of the classical action functional obtained as the
              limit $\l_0\ra0$, e.g.,
               $\p(t,x)=|x|^{-s/3}\s \exp[-t^2-x^2]$. From a path integral viewpoint, this domain modification leads to a different
                set of paths than in the free path integral. Moreover, the spectrum of the relevant quantum Hamiltonian is a uniform
                spectrum $(2\hbar\s m)\s k$,
              where $k\in\{0,1,2,3,\cdots\}$, which correctly reflects a vanishing zero-point energy \cite{BCQ}.

\section{Additional EQ  and/or Affine Variable \\Applications}
EQ has also been applied to other problems, including covariant scalar fields \cite{E3,E1,E2,PP,JK1}, which also uses reducible affine fields,
 and quantum gravity \cite{E3,G1,G2,G3}, which exploits irreducible affine fields. The use of affine variables to discuss idealized
 cosmological models \cite{Y1,WW} has led to gravitational bounces rather than an initial singularity of the universe. Further use of
 affine vbles as applied to idealized gravitational models has been given in various papers, e.g., \cite{T1,T2,T3}, and
 references therein.
\section*{Acknowledgements}The author thanks Ewa Czuchry for useful suggestions.

\end{document}